\documentclass[12pt]{article}
\usepackage{jheppub}
\usepackage{amsmath,amssymb,amsbsy}
\usepackage{graphicx,color}
\usepackage[dvipsnames]{xcolor}
\usepackage{cancel,slashed,mathrsfs}
\usepackage{hyperref}
\usepackage{framed}

\newcommand{\be}{\begin{eqnarray}}
\newcommand{\ee}{\end{eqnarray}}
\newcommand{\nn}{\nonumber}
\newcommand{\Tr}{\mathrm{Tr}}
\newcommand{\NN}{\mathcal{N}}
\newcommand{\CA}{{\mathcal{A}}}
\newcommand{\tO}{{\widetilde{O}}}
\newcommand{\GF}[1]{\Gamma\!\left(#1\right)}
\newcommand{\sfrac}[2]{\mbox{$\frac{#1}{#2}$}}
\newcommand{\LL}{{\mathcal L}}
\def\al{\alpha}
\def\eps{\epsilon}

\title{Two-point functions in $4-2\,\varepsilon$ dimensions from localization}

\author[a]{Alessandro Georgoudis,}
\author[b,c,d]{Joseph A. Minahan,}
\author[e]{Anton Nedelin,}
\author[a]{Congkao Wen}

\affiliation[a]{Centre for Theoretical Physics and Astronomy, Department of Physics and Astronomy,\\
  Queen Mary University of London, Mile End Road, London E1~4NS, United Kingdom}
\affiliation[b]{Department of Physics and Astronomy, Uppsala University,\\
  Box 524, SE-751~20 Uppsala, Sweden}
  \affiliation[c]{Centre for Geometry and Physics, Uppsala University\\
  Box 480, SE-751~06 Uppsala, Sweden}
\affiliation[d]{Center for Theoretical Physics -- a Leinweber Institute,\\ Massachusetts Institute of Technology, Cambridge, MA 02139, USA}
\affiliation[e]{Department of Mathematics, King's College London,\\
  London, WC2R~2LS, United Kingdom}

\preprint{{\raggedleft UUITP-12/26\\ MIT-CTP/6055\\ QMUL-PH-26-23 \par}}

\abstract{%
We study two-point functions of half-BPS operators in maximally supersymmetric Yang-Mills theory continued to $d=4-2\,\varepsilon$ dimensions.  Using supersymmetric localization on $S^d$, we derive perturbative matrix-model expressions for the $\varepsilon$-expansion of these correlators and obtain all-loop results at leading order in $\varepsilon$ in the planar limit, with extensions to finite-$N$ corrections and higher-charge operators.  We compare the localization results with direct perturbative computations in flat space. At order $\varepsilon$ the two descriptions agree perfectly, while at higher orders our construction fails to reproduce the perturbative data due to the breaking of conformal symmetry away from four dimensions. Nevertheless, in the case of the dimension-two operator we conjecture an all-loop formula at order $\varepsilon^2$ by exploiting the precise form of the mismatch.
}

\begin{document}
\maketitle

\section{Introduction}
$\NN=4$ super Yang-Mills (SYM) is an extremely useful laboratory for investigating various perturbative and nonperturbative behavior in quantum field theories.  The large amount of supersymmetry allows us to use various tools to greatly simplify computations.  One of the most important of these is localization \cite{Pestun:2016zxk}, where certain calculations can be reduced to evaluating matrix integrals.  The main requirement is that the observable one is computing is protected by supersymmetry.  This can be the expectation values of certain Wilson loops, or for the purposes of this paper, integrated correlators.

Localization is not just restricted to four dimensions, it can also be applied to maximally supersymmetric gauge theories in any dimension, including non-integer dimensions \cite{Minahan:2015any,Gorantis:2017vzz}.  In particular, at large $N$, it is possible to find the partition function and the expectation values of equatorial Wilson loops for Euclidean theories on the $d$-dimensional sphere $S^d$. For the relevance of this paper, we will extract two-point correlators of BPS operators from the partition function on  $S^d$.

Recently, Bianchi has considered two-point functions of BPS operators in $d=4-2\,\varepsilon$ dimensions in $\mathcal{N}=4$ SYM \cite{Bianchi:2021ohn,Bianchi:2023cbc,Bianchi:2023llc} and also in $d=3-2\,\varepsilon$ in ABJM \cite{Bianchi:2024nah,Bianchi:2025sjc}. By moving to $4-2\,\varepsilon$ dimensions the conformal symmetry is broken and the correlator contains extra terms that depend on $\varepsilon$.  In $d=4$, two- and three-point functions of these half-BPS operators are protected by supersymmetry and are tree-level exact \cite{Lee:1998bxa, Baggio:2012rr}. However, away from four dimensions, the correlation functions receive non-trivial quantum corrections and become interesting functions of the coupling.  Furthermore, the gauge coupling is no longer dimensionless, so a scale $\mu$ must be included to define a dimensionless 't Hooft constant.  The perturbative corrections for two-point functions in $\mathcal{N}=4$ SYM in $4-2\,\varepsilon$ dimensions that will be relevant for this paper have been obtained up to three-loop order \cite{Bianchi:2021ohn,Bianchi:2023cbc,Bianchi:2023llc}, using results from Feynman diagram computations \cite{Baikov:2010hf,Lee:2011jt,Georgoudis:2018olj,Georgoudis:2021onj}, and a conjectured form for the all-loop result at first order in $\varepsilon$.

Using localization, we find an all-loop result for these correlators.  In this case, the scale is $\mu=1/r$, where $r$ is the radius of $S^d$.  Our results match with those in \cite{Bianchi:2021ohn,Bianchi:2023cbc,Bianchi:2023llc} from direct Feynman diagram calculations to first order in $\varepsilon$, confirming the conjecture for the all loop result. At the next order we find discrepancies starting at two loops. This is likely due to scheme-dependent effects, since introducing a scale by putting the theory on the sphere is not the only way to produce a dimensionless 't Hooft parameter. In the special case of the two-point function for dimension-two operator, which sits in the stress-tensor supermultiplet, we conjecture the precise form of the discrepancy at order $\varepsilon^2$, thereby obtaining an all-loop expression for the two-point function at this order.

The rest of the paper is organized as follows.  In section \ref{sec:loc:rev} we briefly review the localization of maximal SYM on $S^d$ in non-integer dimension and recall the resulting matrix model. In section \ref{sec:two:point:loc} we explain how two-point functions of half-BPS operators are extracted from this matrix model, including the Gram-Schmidt orthogonalization needed to compare with flat-space operators. We also comment on the regime of validity of the localization results when compared with flat-space. In section \ref{sec:two:point} we work out explicit examples, derive the all-loop planar result at order $\varepsilon$, and discuss its relation to integrated four-point correlators. We then compare with available perturbative data in flat space in section \ref{sec:comparisonwpd} and give a conjecture for the order-$\varepsilon^2$ mismatch and the associated all-loop result. Finally in section \ref{sec:conclusion} we discuss possible future directions and interesting properties of our results. Additional finite-$N$ examples for higher-dimensional operators are collected in appendix \ref{app:higher}.

\section{Review of localization in non-integer dimensions}
\label{sec:loc:rev}

In this section we review localization of maximally supersymmetric gauge theories in dimension $d$, where $d$ is not necessarily an integer \cite{Minahan:2015any,Gorantis:2017vzz}.  We will be particularly concerned with $d=4-2\,\varepsilon$.  In any dimension, there are 16 total supersymmetries.  The procedure follows Pestun's method of reducing $SU(N)$ super Yang-Mills in ten dimensions to four \cite{Pestun:2007rz}.  The resulting on-shell action was originally found in \cite{Blau:2000xg}.  The off-shell version necessary for localization was derived in \cite{Minahan:2015jta}.

The 10-dimensional Lagrangian is given by \cite{Brink:1976bc}
 \be\label{LL}
\LL= \frac{1}{g_{10}^2}\Tr\left(\sfrac12F_{MN}F^{MN}-\Psi\slashed{D}\Psi\right)\,,
 \ee
where the $\Psi_\alpha$ are Majorana-Weyl fermions that transform in the adjoint representation of the gauge group.  This action is invariant under the supersymmetry transformations
\be\label{susy}
 \delta_\eps A_M&=&\eps_\al\,\Gamma_M^{\al\beta}\Psi_\beta\,,\nn\\
  \delta_\eps \Psi_\al&=&\sfrac12 {{\Gamma^{MN}}_\al}^\beta F_{MN}\,\eps_\beta\,,
 \ee
where we use the convention that $\eps_\al$ is a bosonic ten-dimensional spinor.

This theory can be dimensionally reduced to $d$ Euclidean dimensions by assigning $d$ of the 10-dimensional gauge fields to $A_\mu$, $\mu=1,\dots d$, and the remaining $10-d$ to $\phi_I$, $I=0,\dots 9-d$.  The $\phi_0$ field descends from the 10-dimensional time-like direction and hence has the wrong-sign kinetic term.  The reduced theory thus has an $SO(1,9-d)$ $R$-symmetry.  The reduced field strengths are
\be
 F_{\mu I}&=&[D_\mu,\phi_I] \,, \nn\\
 F_{IJ}&=&[\phi_I,\phi_J]\,,
 \ee
 while the reduced Yang-Mills coupling is $g_{\rm YM}^2=g_{10}^2/V_{10-d}$, where $V_{10-d}$ is the volume of the reduced space. 

Putting the theory on the sphere $S^d$ requires extra terms in the supersymmetry transformations.  The result is that \eqref{susy} is modified to
 \be\label{susysp}
 \delta_\eps A_M&=&\eps\,\Gamma_M\Psi\,,\nn\\
  \delta_\eps \Psi&=&\sfrac12 \Gamma^{MN}F_{MN}\eps+\frac{\alpha_I}{2}\Gamma^{\mu I}\phi_I\nabla_\mu\,\eps\,.
 \ee
Here the index $I$ is summed over and the constants $\alpha_I$ are given by
 \be\label{alrel}
\alpha_I&=&\frac{4(d-3)}{d}\,,\qquad I=8,9,0\, ; \nn\\
\alpha_I&=&\frac{4}{d}\,,\qquad I=d+1,\dots 7\, .
\ee
Furthermore, $\eps_\al$ is no longer constant, but instead satisfies
\be\label{KS}
\nabla_\mu\eps=\beta\,\tilde\Gamma_\mu\Lambda\, \eps\,,
\ee
where $\beta=1/(2r)$ with $r$ being the radius of $S^d$ and $\Lambda=\Gamma^8\Gamma^9\Gamma^0$.  It is clear that the $R$-symmetry is now broken to $SO(6-d)\times SO(1,2)$, except in $d=4$, where it remains $SO(1,5)$.  With these changes, the Lagrangian is now
\be\label{Lss}
\LL_{ss}&=&\frac{1}{g_{\rm YM}^2}\Tr\Bigg(\sfrac12F_{MN}F^{MN}-\Psi\slashed{D}\Psi+\frac{(d-4)}{2r}\Psi\Lambda\Psi+\frac{2(d-3)}{r^2} \phi^A\phi_A+\frac{(d-2)}{r^2}\phi^i\phi_i\nn\\
&&\qquad\qquad\qquad\qquad -\frac{2}{3r}(d-4)[\phi^A,\phi^B]\phi^C\varepsilon_{ABC}\Bigg)\,,
\ee
where we have split the scalars into two types, those with indices $A=8,9,0$ and those with indices $i=d+1,\dots 7$.

In order to localize, it is necessary to consider off-shell supersymmetry transformations.  For general $d$ the transformations are a natural generalization of the $d=4$ case in \cite{Pestun:2007rz}, given by
\be\label{susyos}
 \delta_\eps A_M&=&\eps\,\Gamma_M\Psi\,,\nn\\
  \delta_\eps \Psi&=&\sfrac12 \Gamma^{MN}F_{MN}\eps+\frac{\alpha_I}{2}\Gamma^{\mu I}\phi_I\nabla_\mu\,\eps+K^m\nu_m\, ,\nn\\
\delta_\eps K^m&=&-\nu^m\slashed{D}\Psi+\Delta K^m\,,
 \ee
where the $\nu^m$ are pure-spinors chosen to satisfy
 \be\label{psprop}
 \eps\Gamma^M\nu_m&=&0 \, , \nn\\
 \nu_m\Gamma^M\nu_n&=& \delta_{nm}\eps\Gamma^M\eps=\delta_{nm}v^M\,.
 \ee
Here $v^m$ is a vector field with at least some of its components along the sphere and the rest along the reduced dimensions.  The auxiliary field $K^m$ contributes the term
\be
 \LL_{aux}=-\frac{1}{g_{\rm YM}^2}\,\Tr K^mK_m\, .
 \ee

If we omit the instanton contributions (in particular they are suppressed in the 't Hooft large-$N$ expansion), then localization forces the following condition,
\be\label{fpeq}
\nabla_\mu\phi^I\nabla^\mu\phi^I+(K^m+2\beta(d-3)\phi_0(\nu_m\Lambda\eps))^2+\frac{\beta^2d^2}{4}\sum_{I\ne 0}(\al_I)^2\phi^I\phi^I
=0\, ,\ee
where we further assumed that $v^0=1$, $v^{8,9}=0$.  The fixed point locus is then
\be\label{fpl}
K^m=-2\beta(d-3)\phi_0(\nu^m\Lambda\eps)\,,\qquad\nabla_\mu\phi_0=0\,,\qquad \phi_J=0\ \ {J\ne0}\,,
\ee
which when substituted into the action gives  
\be\label{LLfp}
S_{fp}=+\frac{V_d}{g_{\rm YM}^2}\frac{(d-1)(d-3)}{r^2}\Tr(\phi_0\, \phi_0)=\frac{8\pi^{\frac{d+1}{2}}r^{d-4}}{g_{\rm YM}^2\Gamma\left(\frac{d-3}{2}\right)}\Tr\,\sigma^2\,,
\ee
where $V_d$ is the volume of $S^d$ and $\sigma$ is the dimensionless field $\sigma=r\phi_0$.

On top of this, we need to find the contributions of the Gaussian fluctuations about the fixed points.  These were computed in \cite{Minahan:2015any,Gorantis:2017vzz} for general $d$, where it was found that
\be\label{gauss}
Z_{gauss}(\sigma)=\prod_{\rho>0
}\frac{1}{\langle\rho,\sigma\rangle^2}\prod_{n=0}^\infty\left(\frac{n^2+\langle\rho,\sigma\rangle^2}{(n+d-3)^2+\langle\rho,\sigma\rangle^2}\right)^{\frac{\Gamma(n+d-3)}{\Gamma(n+1)\Gamma(d-3)}}\, ,
\ee
where $\rho$ runs over the positive roots of $SU(N)$.  The $\sigma$ can be diagonalized, which leads to a Vandermonde term that cancels the first factor in \eqref{gauss}.  Hence the final result for the partition function is
\be
Z = \int\prod_{i=1}^{N-1} d\phi_i \prod\limits_{i>j}\prod\limits_{n=0}^{\infty}
\left(\frac{n^2 + \phi_{ij}^2}{(n+d-3)^2+ \phi_{ij}^2} \right)^{\frac{\GF{n+d-3}}{\GF{n+1}\GF{d-3}}}
e^{-\frac{8\pi^{\frac{d+1}{2}}r^{d-4}}{g_{\rm YM}^2\GF{\frac{d-3}{2}}}\sum\limits_{i=1}^N\phi_i^2}\,,
\label{matrix:model:N=4}
\ee
where $\phi_i$ are the eigenvalues for $\sigma$ and $\phi_{ij} \equiv (\phi_i - \phi_j)$. 

In the large-$N$ limit we can solve the system by saddle-point, where we find the eigenvalue equation
\be
\frac{16\pi^{\frac{d+1}{2}}r^{d-4}}{g_{\rm YM}^2\Gamma\left(\frac{d-3}{2}\right)}\,\phi_i= \sum_{j\neq i}G(\phi_{ij})\,,
\label{mm:saddle}
\ee
where the function $G(x)$ is given by
\begin{align}
G(x) ={}& -i\GF{4-d}\left[
\frac{\GF{-ix }}{\GF{4-d-ix}}
-\frac{\GF{ix }}{\GF{4-d+ix}}
-\frac{\GF{d-3-ix }}{\GF{1-ix}}
\right. \notag\\
&\left.
\hspace{32mm}
+\frac{\GF{d-3+ix }}{\GF{1+ix}}
\right]\,.
\label{G:function}
\end{align}
We further define a dimensionless 't Hooft parameter
\be
\lambda = \frac{g_{\rm YM}^2 N\GF{\frac{d-3}{2}}}{16\pi^{\frac{d+1}{2}}r^{d-4}}\,,
\label{tHooft:coupling}
\ee
which simplifies the form of  \eqref{mm:saddle}.  With this definition, we can express the partition function as
\be
Z = \int\prod_{i=1}^{N}d\phi_i \, \delta\left(\sum_{i=1}^N\phi_i\right)  e^{-\frac{N}{2\lambda}\sum\limits_{i=1}^N \phi_i^2 + \sum\limits_{i\neq j}H(\phi_{ij})}\,,
\label{matrix:model:final}
\ee
where the $\delta$-function is inserted to enforce the trace condition for the $SU(N)$ gauge group.  We have introduced the function $H(x)$, which is related to $G(x)$ in \eqref{G:function} by
\be
\partial_x H(x) \equiv \frac{1}{2} \, G(x)\, ,
\label{H:function}
\ee
where the factor of $1/2$ is to account for the double counting.

\section{Two-point correlators from localization}
\label{sec:two:point:loc}

We now explain how the localized matrix model is used to compute two-point functions of half-BPS operators in $\mathcal{N}=4$ SYM continued to $d=4-2\,\varepsilon$ dimensions.  In the field theory, these operators are built as traces over fundamental scalar fields.  In the localization matrix model they are represented by traces of powers of the matrix-model scalar $\phi$.  For example, a single-trace operator with dimension $p$, $O_p$, is mapped to $\Tr \phi^p:=\sum_i \phi_i^p$, and similarly for multi-trace operators.\footnote{In principle, it is not clear whether inserting higher-dimensional operators would not affect the localized partition function given in \eqref{gauss}, which we know is the case only for $\varepsilon=0$. This subtlety is relevant for studying the correlators beyond order $\mathcal{O}(\varepsilon^1)$. As we will discuss shortly, the localization computation will produce the flat-space results at order $\mathcal{O}(\varepsilon^1)$ but not at higher order, therefore we will not be concerned with this issue. We will come back to this point when we compare the localization results with the flat-space computations.} The two-point functions of interest can then be expressed as correlators of $\Tr \phi^p$ in the matrix model, for example,
\be
\langle O_2 \, O_2\rangle_S := \langle \Tr \phi^2\,  \Tr\phi^2 \rangle\,,
\label{O2O2:mm:def}
\ee
where the subscript $S$ emphasizes that the correlators are on $S^d$. A general matrix model correlator has the form
\be
\left\langle \prod_{j=1}^M\Tr\phi^{m_j}\right\rangle = \frac{1}{Z}\int\prod_{i=1}^{N}d\phi_i \, \delta\left(\sum_{i=1}^N\phi_i\right) \prod_{j=1}^M\left(\sum\limits_{i=1}^{N} \phi_i ^{m_j}\right)
e^{-\frac{N}{2\lambda}\sum\limits_{i=1}^N \phi_i^2 + \sum\limits_{j\neq i}H(\phi_{ij})}\,.
\label{mm:ev}
\ee

As emphasized above, the matrix model correlators are defined on $S^d$, so they must be mapped back to $R^d$. This is standard for conformal theories through a Gram-Schmidt orthogonalization procedure~\cite{Gerchkovitz:2016gxx}, which we will review shortly.  However, in our case the deviation from $d=4$ breaks conformal symmetry, so we must take into account the difference between correlators on $S^d$ and those on $R^d$.
 The difference of correlators on $S^d$ and $R^d$ takes the following form, 
\begin{equation}\label{eq:mismatchsft}
\langle O (x_1) \, O (x_2) \rangle_{S} - \langle O(x_1) \, O(x_2) \rangle_{R}  \propto   \varepsilon \int d^4 y\, \sigma(y) \langle O (x_1) O (x_2)  \mathcal{L} (y) \rangle_{R}  + \mathcal{O}(\varepsilon^2) \, ,
\end{equation}
where $\mathcal{L}$ is the Lagrangian of $\mathcal{N}=4$ SYM and $\sigma(y)$ is the Weyl rescaling factor, relating the sphere metric to the flat metric, 
\begin{equation}
g^{(S)}_{\mu \nu} (y) =  e^{2\sigma(y)} \delta_{\mu\nu} \, . 
\end{equation}
For half-BPS operators $O$, supersymmetric non-renormalization gives $\langle O O  \mathcal{L}  \rangle_{R}=0$~\cite{Baggio:2012rr}. Therefore even though in $d=4-2\,\varepsilon$ dimensions the theory is not conformal, we expect that the localization computation of the correlators on a sphere will still match the direct field theory calculation on flat space at order $\mathcal{O}(\varepsilon^1)$ to all orders in the coupling. As we will discuss further, this will be confirmed through explicit Feynman diagram computation.  The regime of validity identified here is analogous to that encountered in recent studies of correlators and Wilson lines in non-conformal theories using supersymmetric localization~\cite{Billo:2024hvf, Billo:2024fst, Billo:2025sxr}, where similar constraints on the applicability of localization techniques arise when deforming away from the conformal point.

\subsection{\texorpdfstring{$\varepsilon$-expansion of the sphere partition function}{epsilon-expansion of sphere partition function}}

Since we are interested in the small-$\varepsilon$ and small-coupling expansion of the correlators, we perform the following operations on the sphere partition function.  First, we rescale the Coulomb branch parameters $\phi_i \to\sqrt{2\lambda/N}\phi_i$ so that the $\lambda$ dependence is
removed from the exponential factor of the matrix model \eqref{matrix:model:final}, leading to
\be
Z = \left(\frac{2\lambda}{N}\right)^{\frac{N-1}{2}}\int \prod\limits_{i=1}^{N}d\phi_i  \, \delta\left(\sum_{i=1}^N\phi_i\right) e^{-\sum\limits_{i=1}^N\phi_i^2}
e^{\sum\limits_{i\neq j}^{N}H\left(\sqrt{\frac{2\lambda}{N}}\phi_{ij}\right)}\,.
\ee
Now using the explicit expressions in \eqref{G:function} and \eqref{H:function}, we can expand the logarithm of the one-loop determinant of the above matrix model in powers of $\lambda$ and rewrite it in the following form:
\be
H\left(\sqrt{2\lambda/N} \phi_{ij}\right) =\frac{1}{2}\log\left(\frac{2\lambda}{N}\right) + \frac{1}{2} \log\phi_{ij}^2 +
\sum\limits_{l>0}\sum\limits_{k>0}F_{k,l}\varepsilon^k\left(\frac{2\lambda}{N}\right)^l\phi_{ij}^{2l}\, , 
\label{H:expansion}
\ee
where $F_{k,l}$ are coefficients of the double expansion of the one-loop determinant in $\varepsilon$ and $\lambda$. These coefficients can be evaluated directly.
First of all, the coefficients at order $\varepsilon^0$ satisfy
\be
F_{0,0} = 1\,,\qquad F_{k,0} = F_{0,l}=0\quad \forall ~l,k>0\, , 
\ee
{\it i.e.}, the expansion starts at $1$ and has no loop corrections at order $\mathcal{O}(\varepsilon^0)$.
At order $\mathcal{O}(\varepsilon^1)$, we  conjecture an all orders in $\lambda$ expansion,  with coefficients
\be
F_{1,l} = (-1)^l\zeta_{2l+1}\frac{2l+1}{l}\, ,
\label{coeffs:e:order}
\ee
which follows from the behavior 
of a large number of terms.
Similarly, at order  $\mathcal{O}(\varepsilon^2)$, we conjecture that the coefficients take the relatively simple form
\begin{equation}\label{coeffs:e2:order}
F_{2,l}=\frac{(-1)^l(2l+1)}{2l}
\left[
(2l+1)\zeta_{2l+2}
+
2\sum_{j=1}^{l-1}
\zeta_{2j+1}\zeta_{2l-2j+1}
\right] \, . 
\end{equation}
Importantly, the $\varepsilon$ expansion of the partition function at the leading order $\mathcal{O}(\varepsilon^1)$ (i.e. $F_{1,l}$), up to a factor of $l$, is identical to the sphere partition function of $\mathcal{N}=2^*$ SYM in the small-mass expansion to the first non-trivial order (i.e. $\mathcal{O}(m^2)$, where $m$ is the mass deformation)~\cite{Russo:2013kea}. This is of particular interest because the small-mass expansion of the $\mathcal{N}=2^*$ partition function directly determines the integrated four-point correlators in $\mathcal{N}=4$ SYM~\cite{Binder:2019jwn}, whose expression is known for a generic rank of the gauge group and any value of the coupling~\cite{Dorigoni:2021bvj, Dorigoni:2021guq}. We will come back to this important point later. 

In order to evaluate correlators in the interacting theory, it is useful to rewrite  $\sum_{i,j}\phi_{ij}^{2l}$ as
\be
\sum\limits_{i,j}\left( \phi_i - \phi_j \right)^{2l} = \sum_{k=0}^{2l}(-1)^k\binom{2l}{k}\Tr\phi^k \, \Tr\phi^{2l-k} \, .
\ee
The matrix model correlators are then written in the form
\be
\langle \Tr\phi^{n_1}\Tr\phi^{n_2}\dots\rangle = \frac{1}{Z}\left(\frac{2\lambda}{N}\right)^{\frac{N^2-1+n_1+n_2+\dots}{2}}
\int d \phi \Tr\phi^{n_1}\Tr\phi^{n_2}\dots e^{-\Tr\phi^2}
\nn\\
\exp\left[\sum\limits_{l>0}\sum\limits_{k>0}F_{k,l}\varepsilon^k\left(\frac{2\lambda}{N}\right)^l
\sum\limits_{m=0}^{2l}(-1)^m\binom{2l}{m}\Tr\phi^m\Tr\phi^{2l-m}\right] \, ,
\label{mm:correlators:gen}
\ee
where the partition function $Z$ is given by the same  expression without the insertion of the $\Tr \phi^{n_i}$ operators. 

\subsection{Gram-Schmidt orthogonalization}

As emphasized above, supersymmetric localization computes correlators on $S^d$, where operators with different dimensions can mix. Such mixing is resolved through a Gram-Schmidt orthogonalization procedure~\cite{Gerchkovitz:2016gxx}, which connects these matrix model correlators to the correlators of operators on the field theory side.
To perform the matching, instead of the operators $O_p \equiv \Tr\phi^p$ in the matrix model, we should consider operators without self-contractions. For this purpose we need to make a given operator $O_p$ orthogonal to all lower-dimensional operators \cite{Gerchkovitz:2016gxx, Billo:2017glv}. Suppose we start with an operator $O_p$ and the operators $\{O_q\}$ form the basis of operators with dimensions lower than or equal to $p-2$. Introducing the matrix of correlators
\be
C_{q_1q_2}\equiv\langle O_{q_1} O_{q_2} \rangle\,,\qquad C^{q_1q_2}\equiv C_{q_1q_2}^{-1}\,,
\label{correlators:matrix}
\ee
we can write an orthogonalized operator 
\be
\tO_p = O_p - \sum\limits_{q_1,q_2}\langle O_pO_{q_1} \rangle C^{q_1q_2}O_{q_2} \, . 
\label{operator:ortho}
\ee
This procedure can be done for any operator, including multi-traces. 
For example, all operators and their orthogonalized versions of weight $p \leq 4$ are given by
\be
&O_2 = \Tr\phi^2\,,\qquad &\tO_2 = \Tr\phi^2 - \langle \Tr\phi^2\rangle\,,
\nn\\
&O_3 = \Tr\phi^3\,, \qquad &\tO_3 = \Tr\phi^3\,,
\nn\\
&O_4 = \Tr\phi^4\,, \qquad  &  \tO_4 = \Tr\phi^4 - \langle \Tr\phi^4\rangle -\tO_2\frac{\langle\tO_2 \Tr\phi^4\rangle}{\langle\tO_2\tO_2\rangle}\,,
\nn\\
&O_{2,2} = (\Tr\phi^2)^2\,,\qquad  &   \tO_{2,2} = (\Tr\phi^2)^2 - \langle \Tr\phi^2 \Tr\phi^2 \rangle - \tO_2\frac{\langle\tO_2 (\Tr\phi^2)^2 \rangle}{\langle \tO_2\tO_2\rangle}\, . 
\label{operator:ortho:examples}
\ee
The flat-space two-point correlators can then be obtained from matrix model by computing 
\begin{align}
  \frac{\langle \tO_p \, \tO_p \rangle~}{\langle \tO_p \, \tO_p \rangle_0} \, . 
\end{align}
We always normalize our correlators with respect to the corresponding tree-level expressions, denoted as $\langle O\, O \rangle_0$,
which can be derived from the usual Gaussian matrix model integral 
\be
Z_0 = \int\prod\limits_{i=1}^{N-1}d\phi_i \prod\limits_{i<j}(\phi_i-\phi_j)^2e^{-\frac{N}{2\lambda}\sum\limits_{i=1}^N \phi_i^2} 
= C_0 \,\lambda^{\frac{N^2-1}{2}}\,,
\label{gaussian:mm}
\ee
where $C_0$ is some constant independent of $\lambda$ that is not relevant for us. 
As an example, applying the construction to the dimension-two operator, we have
\be
\langle \tO_2\, \tO_2\rangle \equiv\frac{4}{N^2}\partial^2_{\lambda^{-1}}\log Z\,.
\label{integrated:correlator:def}
\ee
In the next section, we will perform explicit computations for these two-point functions in the $\varepsilon$-expansion, and compare the expressions with the results obtained from direct Feynman diagrams in flat space. 

\section{Examples of two-point correlators from the matrix model}
\label{sec:two:point}

To explicitly compute the two-point functions using matrix models as we outlined in the previous section, 
we expand the exponent in the second line of \eqref{mm:correlators:gen} in $\varepsilon$ and $\lambda$ and write the perturbative expression for the corresponding 
correlator. Doing so, we will then encounter  Gaussian matrix integrals of the following general form:
\be
t_{n_1,n_2,\dots,n_k}\equiv \langle \Tr \phi^{n_1}\Tr\phi^{n_2}\dots\Tr\phi^{n_k}\rangle_0 \, , 
\ee
where the subscript `$0$' is to indicate that the measure is Gaussian, namely $\exp(-\Tr \phi^2)$.  Such integrals can be done recursively following the ``full algebra method" \cite{Billo:2017glv}. For this we just need $t_0$, $t_1$  and $t_2$ which are easily computed to be
\be
t_0\equiv \langle \Tr 1 \rangle_0 = N,\quad t_1\equiv \langle\Tr\phi\rangle_0=0,\quad  t_2 \equiv \langle \Tr\phi^2\rangle_0 = \Tr\left( T^a T^a\right) = \frac{N^2-1}{2} \,.
\label{t_n:init:cond}
\ee
Performing a single Wick contraction inside an arbitrary correlator $t_{n_1,n_2,\dots}$, leads to the recursion relation \cite{Billo:2017glv}
\be
&t_{n_1,n_2,\dots,n_k} = \frac{1}{2}\sum\limits_{m=0}^{n_1-2}\left(t_{m,n_1-m-2,n_2,\dots,n_k} -\frac{1}{N}t_{n_1-2,n_2,\dots,n_k}\right)
\nn\\
&\hspace{17mm}+ \frac{1}{2}\sum\limits_{j=2}^k n_j \left(t_{n_1+n_j-2,n_2,\dots,n_{j-1},n_{j+1},\dots,n_k} - \frac{1}{N}t_{n_1-1,n_2,\dots,n_j-1,\dots,n_k}\right)\,.
\label{t_n:rec:rel}
\ee
Particularly simple examples of this recursion relation occur when $n_1=0,1,2$, for which we find
\be
t_{0,n_2,\dots,n_k} & = & N t_{n_2,\dots,n_k}\,,\qquad t_{1,n_2,\dots,n_k}=0\,,
\nn\\
t_{2,n_2,\dots,n_k} & = & \frac{N^2-1+n_2+n_3+\dots +n_k}{2}t_{n_2,\dots,n_k}\,.
\ee
Using these Gaussian matrix model correlators, it is now straightforward to evaluate the correlators in
\eqref{mm:correlators:gen} for our interacting matrix model. 

Applying the techniques described above, we can evaluate correlators of the form $\langle \tO^{(i)}_p\, \tO^{(i)}_p\rangle$, where $p$ is the dimension of the operator and the superscript $(i)$ denotes a possible degeneracy. Furthermore, we are interested in the results in the small-$\varepsilon$ expansion, which takes the following general form:
\be
\CA_{p, i} \equiv\frac{\langle\tO^{(i)}_p \, \tO^{(i)}_p \rangle~}{ \langle\tO^{(i)}_p\, \tO^{(i)}_p \rangle_0 } = \sum\limits_{n=0}^{\infty} \varepsilon^n\CA_{p,i}^{(n)}\, .
\label{result:form}
\ee
In the following subsections we consider explicit examples of two-point correlators, and compare the results with the flat-space results through direct Feynman diagram calculations. 

\subsection{\texorpdfstring{$\langle \tO_2\, \tO_2 \rangle $ correlator}{O2 O2 correlator}}

We start with the simplest correlator, $\langle \tO_2\, \tO_2\rangle$. In this case there is no degeneracy and we omit the superscript.   The coefficients of the expansion in \eqref{result:form} are given by
\be
\CA_2^{(1)} & = & -24\lambda\zeta_{3} + 300\lambda^2\zeta_{5} - 3920\lambda^3\zeta_{7} + \frac{7560 \zeta_{9}\left(7N^2+2\right)}{N^2}\lambda^4  - 
\notag \\
&   &\hspace{6cm} \frac{731808 \zeta_{11}\left(N^2+1 \right)}{N^2}\lambda^5 + \mathcal{O}(\lambda^6)\,,
\nn\\
\CA_2^{(2)} & = & -36\lambda\zeta_{4} + \left[750\zeta_{6}+732\zeta_{3}^2\right]\lambda^2 -\left[ 22240 \zeta_{3}\zeta_{5} + 13720 \zeta_{8} \right]\lambda^3 + 
\notag \\
	   &   &  \left[\frac{374220\left(7N^2+2\right)}{11N^2}\zeta_{10} + \frac{120(1291N^2+726)}{N^2}\zeta_{5}^2
	    +\frac{1680\left(203N^2+18 \right)}{N^2}\zeta_{3}\zeta_{7}\right]\lambda^4
\nn\\	  
	    &   & \hspace{11cm}+ \mathcal{O}(\lambda^5)\,,
\nn\\
\CA_2^{(3)} & = & -128 \lambda \zeta_{5} + \lambda^2\left(2196\zeta_{4}\zeta_{3} +3600\zeta_{7} \right) + 
\notag \\
	    &   &  -\frac{1}{3}\lambda^3\left(166800 \zeta_{6}\zeta_{3} + 100080 \zeta_{4}\zeta_{5} + 71776\zeta_{3}^3 + 250880 \zeta_{9} \right)
\nn\\
	   &   & \left[ \frac{5880\left(203N^2+18\right)}{N^2}\zeta_{8}\zeta_{3} + \frac{600\left(1291N^2+726\right)}{N^2}\zeta_{5}\zeta_{6} + \frac{2520\left(203N^2+18\right)}{N^2}\zeta_{4}\zeta_{7} 
	   +\right.
\nn\\
	   &   & \left.\frac{240\left( 5051N^2+726 \right)}{N^2}\zeta_{3}^2\zeta_{5} + \frac{252000\left(7N^2+2\right)}{N^2}\zeta_{11}\right]\lambda^4 + \mathcal{O}(\lambda^5)\,, 
\label{O2O2:expansion:results}
\ee
where we have set $\CA_2^{(0)}= 1$ to normalize it by the tree-level result. In the planar limit, the results simplify to
\begin{align}
\CA_2^{(1)} ={}& -24\lambda\zeta_{3}
+300\lambda^2\zeta_{5}
-3920\lambda^3\zeta_{7}
+52920\lambda^4 \zeta_{9}
-731808\lambda^5 \zeta_{11}
+ \mathcal{O}(\lambda^6) \,,
\notag \\
\CA_2^{(2)} ={}& -36\lambda\zeta_{4}
+ \left[750\zeta_{6}+732\zeta_{3}^2\right]\lambda^2
-\left[22240 \zeta_{3}\zeta_{5} + 13720 \zeta_{8} \right]\lambda^3
\notag\\
&+ \left[238140\zeta_{10}
+154920\zeta_{5}^2
+341040\zeta_{3}\zeta_{7} \right]\lambda^4
+ \mathcal{O}(\lambda^5)\,,
\notag \\
\CA_2^{(3)} ={}& -128 \lambda \zeta_{5}
+ \lambda^2\left(2196\zeta_{4}\zeta_{3} +3600\zeta_{7} \right)
\notag\\
&-\frac{1}{3}\lambda^3\left[
166800 \zeta_{6}\zeta_{3}
+100080 \zeta_{4}\zeta_{5}
+71776\zeta_{3}^3
+250880 \zeta_{9} \right]
\notag\\
&+\left[
1193640\zeta_{3}\zeta_{8}
+774600 \zeta_{5}\zeta_{6}
+1212240\zeta_{3}^2\zeta_{5}
\right. \notag\\
&\left.\hspace{8mm}
+511560\zeta_{4}\zeta_{7}
+1764000\zeta_{11}
\right]\lambda^4
+ \mathcal{O}(\lambda^5)\, .
\label{O2O2:expansion:results:planar}
\end{align}
We note from the above results that the $\langle \tO_2\, \tO_2 \rangle$ correlator receives non-planar corrections only starting at four loops.  Furthermore, it is straightforward to write general expressions for the first terms of the $\varepsilon$ expansion to all loop orders. We will summarize this derivation in section \ref{sec:O2O2:epsilon:order}, including correlators for higher-dimensional operators.

\subsection{\texorpdfstring{$\langle \tO_3 \, \tO_3 \rangle $ correlator}{O3 O3 correlator}}

Now we consider the $\langle \tO_3 \, \tO_3\rangle$ correlator. We note that $O_3$ mixes only with $\Tr\phi$ which vanishes for $SU(N)$. The results for the first three orders in $\varepsilon$ are given by
\begin{align}
\CA_3^{(1)} ={}& -36\lambda\zeta_{3}
+360\lambda^2\zeta_{5}
-3920\lambda^3\zeta_{7}
+22680\frac{2N^2+1}{N^2}\lambda^4\zeta_{9}
+ \mathcal{O}(\lambda^5) \,,
\notag \\
\CA_3^{(2)} ={}& -54\lambda\zeta_{4}
+ \lambda^2\left(900\zeta_{6} +1224\zeta_{3}^2 \right)
- \lambda^3\left(29440 \zeta_{3}\zeta_{5} + 13720 \zeta_{8} \right)
\notag\\
&+ \lambda^4\left[
102060\zeta_{10}\frac{2N^2+1}{N^2}
+360\zeta_{5}^2\frac{476N^2+363}{N^2}
\right. \notag\\
&\left.\hspace{26mm}
+5040\zeta_{3}\zeta_{7}\frac{74N^2+9}{N^2}
\right]
+ \mathcal{O}(\lambda^5)\,,
\notag \\
\CA_3^{(3)} ={}& -192 \lambda \zeta_{5}
+ \lambda^2\left(3672\zeta_{4}\zeta_{3} +4320\zeta_{7} \right)
\notag\\
&-\frac{64}{189}\lambda^3\left[
217350\zeta_{6}\zeta_{3}
+122535\zeta_{3}^3
+130410\zeta_{4}\zeta_{5}
+246960\zeta_{9}
\right]
\notag\\
&+\lambda^4\left[
\frac{7560 \left(74 N^2+9\right) \zeta_{7} \zeta_{4}}{N^2}
+\frac{1800 \left(476 N^2+363\right) \zeta_{5} \zeta_{6}}{N^2}
\right. \notag\\
&\left.\hspace{11mm}
+\frac{17640 \left(74 N^2+9\right) \zeta_{3} \zeta_{8}}{N^2}
+2160\left(\frac{121}{N^2}+780\right) \zeta_{3}^2 \zeta_{5}
\right. \notag\\
&\left.\hspace{11mm}
+\frac{756000 \left(2 N^2+1\right) \zeta_{11}}{N^2}
\right]
+ \mathcal{O}(\lambda^5)\,.
\label{O3O3:expansion:results}
\end{align}
In the planar limit, we have
\be
\CA_3^{(1)} & = & -36\lambda\zeta_{3}+360\lambda^2\zeta_{5}-3920\lambda^3\zeta_{7}+45360\lambda^4\zeta_{9} + \mathcal{O}(\lambda^5) \,,
\notag \\
\CA_3^{(2)} & = & -54 \lambda  \zeta_{4} + 36 \lambda^2\left(25 \zeta_{6}+34 \zeta_{3}^2\right) 
  	     -40 \lambda^3 \left(343 \zeta_{8}+736 \zeta_{3} \zeta_{5}\right)
\nn\\       &   &  +2520 \lambda^4 \left(81 \zeta_{10}+148 \zeta_{3} \zeta_{7}+68 \zeta_{5}^2\right) + \mathcal{O}(\lambda^5)\,, 
\notag \\
\CA_3^{(3)} & = & -192 \lambda \zeta_{5} + \lambda^2\left(3672\zeta_{4}\zeta_{3} +4320\zeta_{7} \right)
\nn\\
&   & -\frac{320}{3} \lambda^3 \left(414 \zeta_{5} \zeta_{4} +690 \zeta_{3} \zeta_{6} +784 \zeta_{9} +389 \zeta_{3}^3 \right)
\nn\\
&   & +720\lambda^4 \left(2340\zeta_{3}^2\zeta_{5} +1190\zeta_{5}\zeta_{6} +777\zeta_{4}\zeta_{7} +1813\zeta_{3}\zeta_{8} +2100\zeta_{11} \right) + \mathcal{O}(\lambda^5)\,. \nn \\
\label{O3O3:expansion:results:planar}
\ee
The results for the two-point correlators of dimension four and five operators are given in  appendix \ref{app:higher}.

\subsection{\texorpdfstring{$\langle \tO_p \, \tO_p \rangle $ correlator at $\varepsilon$ order}{Op Op correlator at epsilon order}}
\label{sec:O2O2:epsilon:order}

In the previous subsection, we evaluated two-point correlators to a fixed number of loops, as summarized in \eqref{O2O2:expansion:results}.  For the order-$\varepsilon$ contribution to the $O_2$ correlator, however, one can do better and derive an all-loop expression in the planar limit.  The simplification comes from the fact that $\langle \tO_2 \tO_2\rangle$ can be obtained by differentiating the partition function, as in \eqref{integrated:correlator:def}, instead of carrying out the Gram-Schmidt procedure explicitly.

We therefore start from the partition function obtained from \eqref{mm:correlators:gen} by removing the operator insertions.  Keeping only terms linear in $\varepsilon$, the exponential of the interaction reduces to a single insertion in the Gaussian matrix model:
\begin{align}
Z =& \left(\frac{2\lambda}{N}\right)^{\frac{N^2-1}{2}}
\int d \phi e^{-\Tr\phi^2}
\left[1 + \sum\limits_{l>0}F_{1,l}\varepsilon\left(\frac{2\lambda}{N}\right)^l
\sum\limits_{m=0}^{2l}\binom{2l}{m}\Tr\phi^m\Tr\phi^{2l-m}\right] 
\nn\\
=&\, Z_0\left(\frac{2\lambda}{N}\right)^{\frac{N^2-1}{2}}
\left[1 +	 \sum\limits_{l>0}F_{1,l}\varepsilon\left(\frac{2\lambda}{N}\right)^l
\sum\limits_{m=0}^{2l}\binom{2l}{m}\langle\Tr\phi^m\Tr\phi^{2l-m}\rangle_0 \right]\, .
\label{mm:partition:eps:1}
\end{align}
In the second line we have written everything in terms of expectation values $\langle\cdot\rangle_0$ in the Gaussian matrix model.  The factor $Z_0$ is independent of $\lambda$ and will drop out of the normalized correlator.  Since the connected two-point function is obtained from $\log Z$, the relevant linearized expression is
\be
\log Z = \frac{N^2-1}{2}\log\lambda + \sum\limits_{l>0}F_{1,l}\varepsilon\left(\frac{2\lambda}{N}\right)^l
\sum\limits_{m=0}^{2l}\binom{2l}{m}\langle\Tr\phi^m\Tr\phi^{2l-m}\rangle_0  + \mathrm{const}\,.
\ee
What remains is the evaluation of the leading large-$N$ contribution to the Gaussian correlator $\langle \Tr\phi^k\Tr\phi^{2l-k}\rangle_0$.  However, at this order the expectation value factorizes,
\be
\langle\Tr\left(\phi^k \right)\Tr\left(\phi^m\right)\rangle_0 = \langle\Tr\left(\phi^k\right)\rangle_0\langle\Tr\left(\phi^m\right)\rangle_0\,,
\ee
and the remaining single-trace expectation values can be represented as integrals over the planar eigenvalue density,
\be
\langle\Tr\left(\phi^k\right)\rangle_0 = 	 N\int d\phi \rho(\phi) \phi^k = \left( 2N \right)^{1+k/2}\frac{\left( 1 + (-1)^{k} \right)\Gamma\left(\frac{1+k}{2}\right)}
{4\sqrt{\pi}\Gamma\left(2+k/2)\right) } \, , 
\label{single:trace:expec}
\ee
where the eigenvalue density $\rho(\phi)$ is the usual Wigner semicircle for the Gaussian model,
\be
\rho(\phi) = \frac{1}{\pi N}\sqrt{2N - \phi^2}\,.
\label{eigval:dens}
\ee
Substituting the factorized correlators into \eqref{mm:partition:eps:1} gives the compact expression for $\log Z$,
\be
\log Z = \frac{N^2}{2}\left[ \log\lambda + 2\sum\limits_{l>0}F_{1,l}\varepsilon\lambda^l \frac{(2l+2)!(2l)!}{l!(l+1)!^2(l+2)!}\right] \, .
\label{log:Z:final}
\ee
Performing the double derivative with respect to $\lambda^{-1}$ prescribed in \eqref{integrated:correlator:def}, and normalizing by the tree-level result $\langle \tO_2\tO_2\rangle_0$, gives 
\be
\frac{\langle \tO_2 \tO_2 \rangle^{~}}{\langle \tO_2 \tO_2 \rangle_0} = 1 + \varepsilon\sum\limits_{l\geq 1}(-1)^l\frac{(2l+2)!^2(l+1)}{(l+1)!^4(l+2)}\zeta_{2l+1}\lambda^l+\mathcal{O}(\varepsilon^2)\,.
\label{two:point:result}
\ee
The extra factor of $(l+1)$ relative to the momentum-space expression of \cite{Bianchi:2023llc} will be explained in the next section, where we account for the Fourier transform between position and momentum space.  The same matrix-model expansion also allows us to extract finite-$N$ corrections.  For example, the first $1/N^2$ correction is
\begin{equation}\label{eq:NonplanrCorrectionl}
   \frac{\langle \tO_2 \tO_2 \rangle^{~}}{\langle \tO_2 \tO_2 \rangle_{0}}\sim \frac{\varepsilon}{N^2}\sum\limits_{l\geq 1} \frac{(-1)^l\,(l-3)(l-2)(l-1)(l+1)(l+4)\,\lambda^l\,\bigl((2l+2)!\bigr)^2\,\zeta(2l+1)}{24\,(l+2)(2l-1)\,\bigl((l+1)!\bigr)^4}\,.
\end{equation}

We next make the connection with the small-mass expansion of $\mathcal{N}=2^*$ SYM more explicit.  As discussed above, the order-$\varepsilon$ result is closely related to the integrated correlator $\langle O_2O_2O_2O_2 \rangle$ studied in the literature.  The two expressions differ by one power of the loop order, which can be generated by acting with $\lambda \tfrac{d}{d\lambda}$.  To exhibit the relation, we substitute an integral representation of the zeta values in \eqref{two:point:result},
\begin{equation}
    \zeta_i=2^{i-1} \int_0^\infty \frac{d\omega}{\sinh(\omega)^2} \frac{\omega^i}{\Gamma(i+1)}  \, ,
\end{equation}
and then exchange the integration and summation.  The resulting series can be resummed into Bessel functions, 
\begin{align}\label{eq:o2o2Bessel}
   \langle \tO_2 \tO_2\rangle_\varepsilon=&\int_0^\infty d\mu(\omega) \sum\limits_{l\geq 1}(-1)^l\frac{(2l+2)!^2(l+1)}{(l+1)!^4(l+2)}\frac{2^{2l}\omega^{2l+1}}{\Gamma(2l+2)}\lambda^l  \nonumber \\ =&   \int_0^\infty d\mu(\omega) 2\,\omega\left(-1 + J_0^2\!\left(u \right) - J_2^2\!\left(u \right)\right)\,,
\end{align}
where we have defined 
$d\mu(\omega)=\frac{d\omega}{\sinh(\omega)^2}$, $u=4 \sqrt{\lambda}\,\omega$ and 
\be
\langle \tO_2 \tO_2\rangle_\varepsilon = \frac{d}{d\varepsilon} \frac{\langle \tO_2 \tO_2 \rangle^{~}}{\langle \tO_2 \tO_2 \rangle_{0}}\bigg|_{\varepsilon=0} \,.
\ee
Acting with $-\tfrac{\lambda}{4}\tfrac{d}{d\lambda}$ on \eqref{eq:o2o2Bessel} then gives the standard integrated correlator of four $O_2$ operators.  The same relation extends to \eqref{eq:NonplanrCorrectionl} and to subsequent orders in $1/N$, and for generic $N$. Thus we can write
\begin{equation} \label{eq:relation}
    -\frac{g_{\rm YM}^2}{4}\frac{d}{dg_{\rm YM}^2}\langle \tO_2 \tO_2\rangle_\varepsilon= \int d\mu_4\, \langle O_2(x_1)O_2(x_2)O_2(x_3)O_2(x_4)\rangle \,,
\end{equation}
where $g^2_{\rm YM}= 16 \pi^2 \lambda/N$ and $d\mu_4$ is the integration measure for the integrated correlator  given in~\cite{Binder:2019jwn}, whose detailed expression is not needed for the following discussion.  

The analogous relation for higher-charge operators is
\begin{equation} \label{eq:relation2}
    -\frac{g_{\rm YM}^2}{2 p}\frac{d}{dg_{\rm YM}^2}\langle \tO^{(i)}_p \tO^{(j)}_p\rangle_\varepsilon= \int d\mu_4\, \langle O^{(i)}_p(x_1) O^{(j)}_p(x_2)O_2(x_3)O_2(x_4)\rangle \,,
\end{equation}
where $O^{(i)}_p$ can be any dimension-$p$ half-BPS operator and the superscript $(i)$ is to denote possible degeneracies. These relations establish precise connections with integrated four-point functions, and are of practical utility in that they allow one to leverage the exact results for integrated correlators available in the literature. 
In particular, we may organize the operators according to the numbers of $O_2$, one may denote them as $O_M^{(i)}O_2^q$, where dimension-$M$ operators $O_M^{(i)}$ form an orthogonal basis and do not contain $O_2$'s (see \cite{Gerchkovitz:2016gxx} for the detailed construction). It was then shown in \cite{Brown:2023cpz} that the integrated correlators organized in this way obey a recursion relation (called Laplace-difference equation) that relates correlators involving $O_M^{(i)}O_2^q$ for different $q$'s for a given $O_M^{(i)}$. Given the relation \eqref{eq:relation2}, such a recursion relation should also apply to $\langle \tO^{(i)}_p \tO^{(j)}_p\rangle_\varepsilon$. 

In the planar limit, we may focus on single-trace operators, denoted as $O_p$; in this case, a general expression is known for the integrated correlator~\cite{Binder:2019jwn}, which allows us to deduce the two-point function in the $\varepsilon$-expansion to be
\begin{align}\label{eq:opopbessel}
\langle \tO_p \tO_p\rangle_\varepsilon ={}& \int_0^\infty d\mu(\omega)\, 2\omega\bigg[(1-p)\left(1 - J_0^2(u)\right) + p\, J_1^2(u)- J_p^2(u) - 2\sum_{k=1}^{p-1} J_k^2(u)\bigg]\, .
\end{align}
This representation is useful because the strong-coupling expansion can be extracted directly from the Bessel form. For general $p$ we find
\begin{align}
\langle \tO_p \tO_p\rangle_\varepsilon
\overset{\lambda\gg1}{=}
&\ (p-1)\frac12\left(
5+4\gamma_E+2\ln\lambda
\right) \\
&\quad + \Delta_p
-\sum_{\ell=0}^{\infty}
\mathcal P_{p,\ell}\,
\frac{
\lambda^{-\frac32-\ell}
\Gamma\!\left(\ell-\frac12\right)
\Gamma\!\left(\ell+\frac32\right)
\Gamma(3+2\ell)\zeta_{3+2\ell}
}
{
4^{2+3\ell}\pi^{4+2\ell}\Gamma(1+\ell)^2
} \, ,  \nonumber
\end{align}
where $\Delta_p =
\frac{1-3p}{2}+\frac1p+2\text{H}_{p-1},$ with
$\text{H}_x=\sum_{n=1}^{x}\frac{1}{n}$ the first harmonic number evaluated for integer $x$ and 
\begin{align}
\mathcal P_{p,\ell}
=&\,
\frac{
(p-1)r_{0,\ell}
+p\,r_{1,\ell}
-r_{p,\ell}
-2\sum_{k=1}^{p-1}r_{k,\ell}
}{
r_{0,\ell}-r_{2,\ell}
}, \\
r_{k,\ell}
=&\,
\frac{\Gamma\!\left(k+\ell+\frac32\right)}
{\Gamma\!\left(k-\ell-\frac12\right)} \,.
\end{align}
The logarithm in the first line is inherited from the strong-coupling expansion of the Bessel integral. It is already present for $p=2$ and persists for the higher-charge operators through the prefactor and the rational functions $\mathcal P_{p,\ell}$.

\subsubsection{\texorpdfstring{Higher $\varepsilon$ orders}{Higher epsilon orders}}
Let us briefly discuss how the same method extends to higher orders in $\varepsilon$. The main difference from the linear analysis is that at order $\varepsilon^2$ there are two kinds of contributions. The first comes from the coefficient $F_{2,l}$ in the one-loop determinant, while the second comes from expanding the exponential of the order-$\varepsilon$ interaction to second order. This is why products of zeta values appear already at this stage.

Focusing on the coefficient \eqref{coeffs:e2:order}, and again taking the planar limit, the contribution to the normalized $O_2$ two-point function at order $\varepsilon^2$ can be organized as 
\begin{align}\label{eq:eps2Position}
& \frac{\langle \tO_2 \tO_2 \rangle^{~}}{\langle \tO_2 \tO_2 \rangle_{0}} \bigg|_{\varepsilon^2}
\sim 
\sum_{l=1}^{\infty}
\frac{(-1)^l (l+1)(2l+1)}{2(l+2)}
\binom{2l+2}{l+1}^{\!2}
\zeta_{2l+2}\,\lambda^l
\\[4pt]
&+\sum_{l=2}^{\infty}\sum_{r=1}^{l-1}
(-1)^l (l+1)\,\zeta_{2r{+}1}\,\zeta_{2l{-}2r{+}1}
\left[
\frac{1}{l+2}\binom{2l+2}{l+1}^{\!2}
+\frac{8l}{(l-r)\,r}\,U_{l,r}
\right] \lambda^l \, , \notag
\end{align}
where
\begin{align} \label{eq:Ulr}
U_{l,r} ={}& \sum_{h=1}^{r}
h\binom{2(l-r)+1}{l-r-h}
\binom{2(l-r)+1}{l-r-h+1}
\binom{2r+1}{r-h}
\binom{2r+1}{r-h+1}\, .
\end{align}
The first line of \eqref{eq:eps2Position} is the direct contribution of $F_{2,l}$. The double sum in the second line is the contribution from two insertions of the order-$\varepsilon$ interaction, with $r$ splitting the loop order between the two insertions. The finite sum $U_{l,r}$ is the combinatorial factor obtained after evaluating the planar Gaussian contractions. This expression is therefore the direct analogue of \eqref{two:point:result} at order $\varepsilon^2$, but it is structurally richer because the matrix-model expansion contains both genuine $F_{2,l}$ terms and products of the lower-order coefficients.

As we will see in the comparison with perturbation theory, this localization result should not by itself be identified with the full flat-space answer at order $\varepsilon^2$. The map between the sphere and flat space receives additional contributions at this order, and these contributions are precisely where the mismatch with the direct perturbative computation appears.

\section{Comparison with perturbation theory}\label{sec:comparisonwpd}
We now compare the localization result with available perturbative computations in flat space. From the discussion around \eqref{eq:mismatchsft}, we expect a direct match at leading order in $\varepsilon$, while higher orders in $\varepsilon$ can receive additional contributions from the sphere-to-flat-space map.

In \cite{Bianchi:2023llc}, the two-point function of the half-BPS dimension-two operator $O_0 = \Tr(XX)$ in $\mathcal{N}=4$ SYM was computed in dimensional regularization, with $d=4-2\,\varepsilon$, up to three loops and to high order in the $\varepsilon$ expansion (here we follow the notations in \cite{Bianchi:2023llc}).  The result is normalized by the leading-order answer as
\begin{equation}
\langle O_0(p)\,O_0(-p)\rangle = 2(N^2-1)\,B_1(p^2)\Bigl(1 + n^{(1)}\lambda + n^{(2)}\lambda^2 + n^{(3)}\lambda^3 + \mathcal{O}(\lambda^4)\Bigr)\,,
\end{equation}
where $B_1(p^2) = \int \frac{d^{4-2\,\varepsilon}k}{(2\pi)^{4-2\,\varepsilon}}\frac{1}{k^2(k-p)^2}$ is the one-loop bubble, with $p^2=1$, and the 't Hooft coupling is defined as in our conventions after absorbing the extra factors of $e^{\gamma_E\varepsilon}$ and $(4\pi)^\varepsilon$. The coefficients $n^{(l)}$ are the loop corrections to be matched against the localization expansion. Explicitly, the perturbative computation gives
\begin{align}
n^{(1)} &= -12\zeta_{3}\,\varepsilon - 18\zeta_{4}\,\varepsilon^2 + \bigl(6\zeta_{2}\zeta_{3} - 84\zeta_{5}\bigr)\varepsilon^3
+ \bigl(64\zeta_{3}^2 - \tfrac{657}{4}\zeta_{6}\bigr)\varepsilon^4 \nonumber\\
&\quad + \bigl(\tfrac{741}{4}\zeta_{3}\zeta_{4} + 42\zeta_{2}\zeta_{5} - 588\zeta_{7}\bigr)\varepsilon^5 + \mathcal{O}(\varepsilon^6)\,, \nonumber \\ 
n^{(2)} &= 100\zeta_{5}\,\varepsilon + \bigl(244\zeta_{3}^2 + 250\zeta_{6}\bigr)\varepsilon^2
+ \bigl(732\zeta_{3}\zeta_{4} - 100\zeta_{2}\zeta_{5} + 1718\zeta_{7}\bigr)\varepsilon^3 \nonumber\\
&\quad + \Bigl(\tfrac{7288}{3}\zeta_{5}\zeta_{3} - \tfrac{1296}{5}\zeta_{5,3} - 244\zeta_{2}\zeta_{3}^2 + \tfrac{179647}{30}\zeta_{8}\Bigr)\varepsilon^4 + \mathcal{O}(\varepsilon^5)\,, \nonumber \\
n^{(3)} &= -980\zeta_{7}\,\varepsilon + \bigl(-5560\zeta_{3}\zeta_{5} - 3430\zeta_{8}\bigr)\varepsilon^2 \nonumber\\
&\quad + \Bigl(-\tfrac{17560}{3}\zeta_{3}^3 - 13900\zeta_{6}\zeta_{3} - 8340\zeta_{4}\zeta_{5} + 1470\zeta_{2}\zeta_{7} - \tfrac{292220}{9}\zeta_{9}\Bigr)\varepsilon^3 + \mathcal{O}(\varepsilon^4)\,. \label{eq:pertupto3l}
\end{align}
These expressions already indicate why the localization matrix model alone cannot be expected to reproduce the full flat-space answer at all orders in $\varepsilon$. For instance, $n^{(2)}$ contains the multiple zeta value $\zeta_{5,3}$ at order $\varepsilon^4$, while the expansion of the localization one-loop determinant in \eqref{H:expansion} generates only the zeta structures displayed above. The leading order in $\varepsilon$ is special, however, and the following closed-form conjecture was given to be
\begin{equation}\label{eq:allloop:conjecture}
\frac{\langle O_0(p)\,O_0(-p)\rangle^{(l)}}{\langle O_0(p)\,O_0(-p)\rangle^{(0)}} = (-1)^l\,\frac{(2l+2)!^2}{(l+2)(l+1)!^4}\,\zeta_{2l+1}\,\varepsilon + \mathcal{O}(\varepsilon^2)\,.
\end{equation}
Our localization result proves this conjecture after accounting for a simple kinematic difference. At first sight, \eqref{two:point:result} differs from \eqref{eq:allloop:conjecture} by a factor of $(\ell+1)$. This factor arises because the localization computation is naturally in position space, like also in \cite{Bianchi:2023cbc}, whereas the perturbative computation of \eqref{eq:pertupto3l} is in momentum space. The two are related by the Fourier transform
\begin{equation}\label{eq:Fourier}
	\text{FT}\left[\frac{1}{x^{2 \ell}}\right]
	= \int \frac{d^d x}{\pi^{d/2}} \frac{e^{2 i p \cdot x}}{x^{2 \ell}}
	=
	\frac{\Gamma(d/2-\ell)}{\Gamma(\ell) }
	\frac{1}{(p^{2})^{d/2-\ell}}
	\, . 
\end{equation}
By dimensional analysis of the results in \cite{Bianchi:2023cbc,Bianchi:2023llc}, the momentum-space contribution at loop order $\ell$ scales as $p^{\varepsilon(1+\ell)}$. To compare normalized correlators, one must Fourier transform both the loop correction and the tree-level result. Expanding the resulting ratio at small $\varepsilon$ gives
\begin{equation}\label{eq:FourierMomToPos}
  \frac{\langle O_0(x_1)\,O_0(x_2)\rangle^{(l)}}{\langle O_0(x_1)\,O_0(x_2)\rangle^{(0)}} 
  = \text{FT}\!\left[\frac{\langle O_0(p)\,O_0(-p)\rangle^{(l)}}{\langle O_0(p)\,O_0(-p)\rangle^{(0)}}\right]
  = \frac{\langle O_0(p)\,O_0(-p)\rangle^{(l)}}{\langle O_0(p)\,O_0(-p)\rangle^{(0)}}
  \bigl(\ell+1+\ell(\ell+1)\varepsilon+\ldots\bigr)\,.
\end{equation}
This accounts for the apparent missing factor and proves the leading-$\varepsilon$ result obtained in \cite{Bianchi:2023llc}. The same comparison can be extended to more general operators of the type studied in section \ref{sec:O2O2:epsilon:order}. Perturbative data for these operators are also available\footnote{We thank Marco Bianchi for sharing his results with us prior to publication.} \cite{BianchiPData}, and the matching at order $\varepsilon$ works for all available cases. This suggests that the construction is valid beyond the planar limit and beyond the simplest dimension-two operator.

\subsection{\texorpdfstring{A conjecture for $\varepsilon^2$}{A conjecture for epsilon squared}}

It would be interesting to extend this comparison beyond the leading order in
$\varepsilon$. At order $\varepsilon^2$, a direct comparison between the localization answer and the perturbative flat-space result reveals a non-trivial mismatch. The results obtained from localization miss the odd $\zeta_i$ contribution which is present in the result of \cite{Bianchi:2023llc}, once transformed back into
position space, 
\begin{align}
\frac{\langle O_0(x_1)\,O_0(x_2)\rangle^{~~~}}{\langle O_0(x_1)\,O_0(x_2)\rangle^{(0)}}\bigg|_{\varepsilon^2}
={}& \bigl(\textcolor{red}{24\zeta_{3}} - 36\zeta_{4}\bigr)\lambda
+ \bigl(732\zeta_{3}^2 \, \textcolor{red}{-\, 600\zeta_{5}} + 750\zeta_{6}\bigr)\lambda^2 \notag\\
&+ \bigl(-22240\,\zeta_{3}\zeta_{5} \,\textcolor{red}{+ \, 11760\,\zeta_{7}} - 13720\zeta_{8}\bigr)\lambda^3
+ \mathcal{O}(\lambda^4)\,.
\end{align}
In the above expression, we have highlighted these extra terms. 
For the operator $O_2$, this mismatch appears to be captured by the same
protected integrated four-point functional studied in
\cite{Binder:2019jwn,Dorigoni:2021bvj,Chester:2025kvw}. The
order-$\varepsilon^2$ analogue of \eqref{eq:mismatchsft}  contains a contribution
of the schematic form
\begin{align}
    \label{eq:eps2mismatch}
& \langle O (x_1) O (x_2) \rangle_{S}
-
\langle \mathcal{O}(x_1) \mathcal{O}(x_2)  \rangle_{R} \\
 \sim &\, 
\varepsilon^2
\left( \int \! d^4 y\, d^4 z\,
\sigma(y)\sigma(z)\;
\langle \mathcal{O}(x_1)\mathcal{O}(x_2)\mathcal{L}(y)\mathcal{L}(z)\rangle_{R} + \cdots \right) \, , \nonumber
\end{align}
where  the ellipses represent other contributions at order $\varepsilon^2$, including terms such as $\varepsilon \int d^{4-2\,\varepsilon} y\, \sigma(y) \langle O (x_1) O (x_2)  \mathcal{L} (y) \rangle_{R^d, d=4-2\,\varepsilon}$ to next order in $\varepsilon$. 
Since the Lagrangian belongs to the same stress-tensor multiplet as $O_2$, one
may expect this double insertion to be related, through supersymmetric Ward
identities and integrations by part, to an integrated correlator of four bottom components,
$\langle O_2 O_2 O_2 O_2\rangle$, analogous to the derivation of \cite{Binder:2019jwn}.

This relation is suggestive rather than derived. The precise implementation of
this idea is subtle: the weighted insertions, possible integrations by parts,
and contact terms have not been fully controlled here. In particular, the naive factor $\sigma(y)\sigma(z)$ should not be identified directly with the standard integrated-correlator measure and one should include other contributions we have not made explicitly. Nevertheless, the explicit perturbative comparison suggests that, in the scheme relevant for the localization computation, the order-$\varepsilon^2$ mismatch for $O_2$ is captured by the same protected integrated four-point function $\langle O_2 O_2 O_2 O_2\rangle$, which we know to all orders as we discussed previously. This motivates the following all-loop conjecture for the planar contribution at order $\varepsilon^2$ of \cite{Bianchi:2023llc},
\begin{align}\label{eq:eps2Momentum}
& \frac{\langle O_0(p)\,O_0(-p)\rangle^{~~~}}
{\langle O_0(p)\,O_0(-p)\rangle^{(0)}}\bigg|_{\varepsilon^2}
=
 \sum_{l=1}^{\infty}
\frac{(-1)^l (2l+1)}{2(l+2)}
\binom{2l+2}{l+1}^{\!2}
\zeta_{2l+2}\,\lambda^l   \\[6pt]
& +\sum_{l=2}^{\infty}\sum_{r=1}^{l-1}
(-1)^l \,\zeta_{2r+1}\,\zeta_{2l-2r+1}
\left[
\frac{1}{l+2}\binom{2l+2}{l+1}^{\!2}
+\frac{8l}{(l-r)\,r}\,U_{l,r}
\right] \lambda^l  \, , \notag
\end{align}
where $U_{l,r}$ is given in \eqref{eq:Ulr}. 

Extending this logic to higher-dimensional operators $O_p$ appears to be more subtle. The naive generalisation of \eqref{eq:eps2mismatch} extending simply to the integrated correlator $\langle O_p O_p O_2 O_2\rangle$ does not agree with results from a direct Feynman diagram calculation \cite{BianchiPData}, although the discrepancy appears to be minimal\footnote{We note the difference at two loops only appears in the coefficient of $\zeta_3^2$ for all the examples we have considered, including those in appendix \ref{app:higher}.}. As noted earlier, it remains unclear whether the localized partition function on $S^d$ is unaffected by the inclusion of higher-dimensional operators. If not, such corrections could provide a natural explanation for the observed discrepancy when extending the analysis to operators of higher charge.

\section{Conclusion and outlook}\label{sec:conclusion}

In this paper we studied two-point functions of half-BPS operators in maximally supersymmetric Yang-Mills theory continued to $d=4-2\,\varepsilon$ dimensions. The main tool is supersymmetric localization on $S^d$, which reduces the problem to a matrix model whose interaction term can be expanded simultaneously in the 't Hooft coupling and in $\varepsilon$. After reviewing the localized matrix model, we explained how the matrix-model operators are related to the flat-space half-BPS operators, including the Gram-Schmidt orthogonalization needed to remove mixing with lower-dimensional operators on the sphere.

We then used this framework to compute explicit normalized two-point functions. For the operators $O_2$ and $O_3$ we obtained weak-coupling expansions through several orders in $\lambda$ and $\varepsilon$, keeping the finite-$N$ dependence where appropriate. The higher-dimensional examples collected in appendix \ref{app:higher} show that the same method applies to multi-trace sectors and to operators with degeneracies. In the planar limit, the order-$\varepsilon$ correction to $\langle O_2O_2\rangle$ can be resummed to all loop orders. We also extracted the first non-planar correction and wrote the leading-$\varepsilon$ result for general charge in a Bessel-function representation.

An important point in the comparison with perturbation theory is that the localization result is naturally formulated in position space, while the direct diagrammatic results of \cite{Bianchi:2023llc} are given in momentum space.  After accounting for the Fourier transform of the normalized correlator, the order-$\varepsilon$ localization result matches the perturbative flat-space result, including the all-loop form conjectured in \cite{Bianchi:2023llc}. The same agreement holds for the additional perturbative data for higher-charge operators. At order $\varepsilon^2$, however, the localization result differs from the flat-space answer. We argued that this mismatch is consistent with the expected difference between correlators on $S^d$ and on $R^d$, and that for $O_2$ it is naturally related to the protected integrated four-point correlator of stress-tensor multiplet operators. This led to an all-loop conjecture for the planar order-$\varepsilon^2$ contribution in momentum space. Our results also open up several interesting future directions, which we will briefly comment on below.

A natural immediate question is to systematically study the discrepancy between the localization and flat-space results at higher orders in the $\varepsilon$-expansion. As discussed above, at order $\varepsilon^2$ the difference appears to admit an interpretation in terms of integrated four-point functions in flat space for $O_2$.  It would be interesting to understand how this extends to higher-dimensional operators, and whether the higher-order $\varepsilon$-expansion of the two-point functions is connected in some way to the second integrated correlator~\cite{Chester:2020dja, Chester:2020vyz, Alday:2023pet} or to other observables in flat space. 

Our results at order $\varepsilon$ are exact and provide all-order expressions in the strong-coupling expansion. It would be interesting to understand these strong-coupling results from a holographic perspective, along the lines of recent holographic studies of correlators in maximally supersymmetric Yang--Mills theories in $d$ dimensions~\cite{Itzhaki:1998dd, Bobev:2019bvq, Bobev:2025idz}. In particular, higher order terms in the $1/\sqrt{\lambda}$ expansion would encode precise stringy corrections in the dual theory, and it would be interesting to study these from the bulk.

Another interesting direction is to explore whether the present small-$\varepsilon$ localization can be used to study supersymmetric versions of the generalized $F$-theorem \cite{Giombi:2014xxa,Fei:2015oha,Giombi:2015haa}. The natural quantity in this context is $\widetilde F=\sin(\pi d/2)\log Z_{S^d}$, which interpolates between the \(a\)-anomaly in even dimensions and the sphere free energy in odd dimensions. Since localization gives direct access to supersymmetric sphere partition functions and protected integrated correlators, it may provide a useful way to test the monotonicity of $\widetilde F$ along supersymmetric renormalization group flows. A complete application would require a careful treatment of the curvature counterterms and of the distinction between sphere and flat-space observables away from the conformal point.  Our work suggests that such an approach might provide perturbative evidence for generalized $F$-theorems in supersymmetric theories.

As we have already emphasized, it is intriguing that the two-point functions in the leading small-$\varepsilon$ expansion are directly related to the integrated four-point functions, where the latter are known exactly for any value of the coupling constant, including non-perturbative instanton effects~\cite{Dorigoni:2021guq, Dorigoni:2021bvj, Paul:2022piq, Paul:2023rka, Brown:2023cpz, Brown:2023why}. It would be of interest to understand whether the relation \eqref{eq:relation2} remains valid when non-perturbative instantons are included; if not, how the relation might be extended, and furthermore to understand the modular properties of the $\varepsilon$-expansion of the two-point functions, in analogy with those of the integrated four-point functions. 

Finally, the intriguing connections between the small-$\varepsilon$ expansion of two-point functions and integrated correlators deserve better understanding also at the perturbative level from direct flat-space field theory calculations. It has been shown~\cite{Wen:2022oky, Brown:2023zbr, Zhang:2024ypu} that integrated correlators can be expressed in terms of periods of $f$-graph conformal Feynman integrals~\cite{Eden:2011we, Eden:2012tu}. It would be of interest to investigate whether the two-point functions studied here admit a similar integral basis, perhaps along the lines of~\cite{He:2025zbz}. Such a basis, if it exists, may also shed light on the graphical structure underlying the $\varepsilon$-expansion, and as in the case of integrated correlators, could allow for efficient calculations at higher loop orders.

\section*{Acknowledgements}

%
We thank Marco Bianchi for insightful  discussions and for sharing the perturbative data before publication. A.~G.~is supported by a Royal Society funding, URF\textbackslash{R}\textbackslash221015. J.~A.~M. is supported in part by the ERC Advanced Grant ``IGTaFT" and by the Swedish Research Council excellence center grant ``Geometry and Physics", 2022-06593. The research of A. N. is supported by STFC Grant 
No. ST/X000753/1. C. W. is supported by a Royal Society University Research Fellowship No. UF160350 and a STFC Consolidated Grant, ST$\backslash$T000686$\backslash$1 ``Amplitudes, strings \& duality".

\appendix

\section{Further examples of correlators with higher-dimensional operators} \label{app:higher}

In this appendix, we provide further results for two-point correlators in the $\varepsilon$-expansion computed from localization. Here we will list results for operators up to dimension five, and similar results have been obtained for higher-dimensional operators.  As we commented in the main text, the results at the leading order (i.e. $\mathcal{O}(\varepsilon^1)$) fully agree with the direct flat-space Feynman diagram calculation~\cite{BianchiPData}, and we would like to stress that the agreement is for generic $N$. 

\subsection{Correlators of dimension-four operators}

There are two dimension four operators: $O^{(1)}_4:=O_4$ and $O^{(2)}_4:= (O_2)^2$. 
Let us denote these different two-point functions as{\allowdisplaybreaks
\begin{align}
   \langle \tO_4\, \tO_4 \rangle = \sum_{i=1}^{\infty}  \mathcal{A}_{4, 1}^{(i)} \, \varepsilon^i \, , \quad
     \langle \tO_4\, (\tO_2)^2 \rangle = \sum_{i=1}^{\infty}  \mathcal{A}_{4, 2}^{(i)} \, \varepsilon^i  \, , \quad 
       \langle (
       \tO_2)^2\, (\tO_2)^2 \rangle = \sum_{i=1}^{\infty}  \mathcal{A}_{4, 3}^{(i)} \, \varepsilon^i \, . 
\end{align}
At order $\varepsilon^1$, we have
\begin{align}
\mathcal{A}_{4, 1}^{(1)} ={}& -48 \zeta_{3} \lambda
+ \frac{240 \left(2 N^4-6 N^2+27\right) \zeta_{5}}{N^4-6 N^2+18}\lambda^2
- \frac{560 \left(10 N^4+3 N^2+108\right) \zeta_{7}}{N^4-6 N^2+18} \lambda^3
\notag\\
&+\frac{3780 (19 N^6 +83 N^4+168N^2+144 ) \zeta_{9}}{N^2(N^4-6N^2+18)} \lambda^4
+ \ldots \,,
\notag\\
\mathcal{A}_{4, 2}^{(1)} ={}& -48 \zeta_{3} \lambda
+ \frac{120 \left(11 N^2-9\right) \zeta_{5}}{2 N^2-3} \lambda^2
- \frac{19600 N^2 \zeta_{7}}{2 N^2-3} \lambda^3
\notag\\
&+\frac{7560 (40N^4 +53N^2-12 ) \zeta_{9}}{N^2(2 N^2-3)} \lambda^4
+ \ldots \,,
\notag\\
\mathcal{A}_{4, 3}^{(1)} ={}& -48 \zeta_{3} \lambda
+ \frac{600 \left(N^2+3\right) \zeta_{5}}{N^2+1}\lambda^2
- \frac{7840 \left(N^2+6\right) \zeta_{7}}{N^2+1} \lambda^3
\notag\\
&+\frac{15120 (7 N^4 +72 N^2+20 ) \zeta_{9}}{N^2(N^2+1)} \lambda^4
+ \ldots \, ; 
\end{align}
and at order $\varepsilon^2$, we obtain
\begin{align}
 \mathcal{A}_{4, 1}^{(2)} =& -\frac{4 \pi^4  }{5} \lambda +\frac{40  \left(378 \left(8
   N^4-42N^2+135\right) \zeta_{3}^2+\pi^6 \left(2
   N^4-6 N^2+27\right)\right)}{63 \left(N^4-6
  N^2+18\right)} \lambda^2 \cr 
   & -\frac{4 
   \left(5400 \left(286 N^4-627 N^2+3672\right) \zeta_{3} \zeta_{5}+7 \pi^8 \left(10 N^4+3
   N^2+108\right)\right)}{135 \left(N^4-6
   N^2+18\right)} \lambda^3 + \ldots \, ,  \cr 
 \mathcal{A}_{4, 2}^{(2)} =& 
   -\frac{4 \pi^4 }{5}\lambda  +\frac{20 \left(1890
   \left(7 N^2-9\right) \zeta_{3}^2+\pi^6 \left(11
   N^2-9\right)\right)}{63 \left(2 N^2-3\right)} \lambda^2  \cr 
   & -\frac{4320 \left(839 N^2-486\right) \zeta_{3} \zeta_{5}+196
   \pi^8 N^2 }{27
   \left(2 N^2-3\right)} \lambda^3 + \ldots  \, ,  \cr 
 \mathcal{A}_{4, 3}^{(2)} =& 
   -\frac{4 \pi^4 }{5}\lambda  +\frac{20 \left(378 \left(17 N^2+27\right) \zeta_{3}^2+5
   \pi^6 \left(N^2+3\right)\right)}{63
   \left(N^2+1\right)} \lambda^2  \cr 
   & -\frac{8 \lambda^3 \left(5400 \left(184 N^2+699\right)
   \zeta_{3} \zeta_{5}+49 \pi^8
   \left(N^2+6\right)\right)}{135 \left(N^2+1\right)} \lambda^3 + \ldots \, .
\end{align}
}

\subsection{Correlators of dimension-five operators}

For dimension five, we have $O_5^{(1)}:=O_5$ and $O_5^{(2)}:=O_3O_2$, and we denote these different two-point functions as
\begin{align}
   \langle \tO_5\tO_5 \rangle = \sum_{i=1}^{\infty}  \mathcal{A}_{5, 1}^{(i)} \, \varepsilon^i \, , \quad
     \langle \tO_5 \tO_{2,3} \rangle = \sum_{i=1}^{\infty}  \mathcal{A}_{5, 2}^{(i)} \, \varepsilon^i  \, , \quad 
       \langle \tO_{2,3} \tO_{2,3} \rangle = \sum_{i=1}^{\infty}  \mathcal{A}_{5, 3}^{(i)} \, \varepsilon^i \, . 
\end{align}
We find 
\begin{align}
   \mathcal{A}_{5, 1}^{(1)} =&\, - 60   \zeta_{3} \lambda +\frac{600  \left(N^4+6N^2+12\right) \zeta_{5}}{N^4+24} \lambda^2 - \frac{1400  \left(5 N^4+57 N^2+48\right) \zeta_{7}}{N^4+24} \lambda^3 \nonumber \\ 
   & +  \frac{12600  \left(7N^6 + 122 N^4+126 N^2+72\right) \zeta_{9}}{N^2(N^4+24)} \lambda^4 + \ldots \, , \cr
    \mathcal{A}_{5, 2}^{(1)} =&\, - 60   \zeta_{3} \lambda + \frac{120  \left(6N^2-5\right) \zeta_{5}}{N^2-2} \lambda^2 - \frac{560  \left(17 N^2+8\right) \zeta_{7}}{N^2-2} \lambda^3 
   \cr &+ \, \frac{3780  \left(35N^4+ 87 N^2+20\right) \zeta_{9}}{N^2(N^2-2)} \lambda^4 + \ldots \, , 
   \cr 
    \mathcal{A}_{5, 3}^{(1)} =&\, - 60   \zeta_{3} \lambda +\frac{60  \left(11N^2+97\right) \zeta_{5}}{N^2+5} \lambda^2 - \frac{7840 \left( N^2+14\right) \zeta_{7}}{N^2+5} \lambda^3 
   \cr 
   & +\, \frac{7560 \left( 13N^4 + 268N^2+133\right) \zeta_{9}}{N^2(N^2+5)} \lambda^4  + \ldots \, ,  
\end{align}
and 
\begin{align}
   \mathcal{A}_{5, 1}^{(2)} = &\, -\pi^4 \lambda +\frac{20  \left(378
   \left(23 N^4+30 N^2+492\right) \zeta_{3}^2+5 \pi^6
   \left(N^4+6 N^2+12\right)\right)}{63
   \left(N^4+24\right)} \lambda^2 \cr 
   &\, -\frac{14 \left(5400
   \left(23 N^4+165 N^2+264\right) \zeta_{3} \zeta_{5}+\pi^8 \left(5 N^4+57 N^2+48\right)\right)}{27
   \left(N^4+24\right)} \lambda^3 + \ldots \, ,  \cr 
      \mathcal{A}_{5, 2}^{(2)} = &\, -\pi^4 \lambda + \frac{20  \left(378 \left(24 N^2-41\right)
   \zeta_{3}^2+\pi^6 \left(6 N^2-5\right)\right)}{63
   \left(N^2-2\right)} \lambda^2 \cr 
   &\, -\frac{28 \left(5400 \left(71 N^2-37\right) \zeta_{3}
   \zeta_{5}+\pi^8 \left(17 N^2+8\right)\right)}{135
   \left(N^2-2\right)} \lambda^3 + \ldots \, , \cr 
      \mathcal{A}_{5, 3}^{(2)} = &\, -\pi^4 \lambda + \frac{10 \left(378 \left(47 N^2+277\right) \zeta_{3}^2+\pi^6 \left(11 N^2+97\right)\right)}{63
   \left(N^2+5\right)} \lambda^2 \cr 
   &\, -\frac{56 \left(1350 \left(127 N^2+1265\right) \zeta_{3}
   \zeta_{5}+7 \pi^8 \left(N^2+14\right)\right)}{135
   \left(N^2+5\right)} \lambda^3 + \ldots \, . 
\end{align}

\bibliographystyle{JHEP}
\bibliography{refsc}

\end{document}